\documentclass[doublecol]{epl2}

\usepackage{epsfig,ams,amsmath}

\title{Nonequilibrium charge dynamics of light-driven  rings threaded by a magnetic flux}

\author{A. S. Moskalenko\inst{1,3}\footnote{Also at: A.F. Ioffe Physico-Technical Institute,
 194021 St. Petersburg, Russia}, A. Matos-Abiague\inst{2} \and \and J. Berakdar\inst{3}}
\shortauthor{A. S. Moskalenko \etal}

\institute{ \inst{1} Max-Planck-Institut f\"{u}r
Mikrostrukturphysik - Weinberg 2, 06120 Halle, Germany\\
  \inst{2} Institute for Theoretical Physics, University of Regensburg - 93040 Regensburg, Germany\\
  \inst{3} Institut f\"ur Physik, Martin-Luther-Universit\"at
 Halle-Wittenberg,  Nanotechnikum-Weinberg, Heinrich-Damerow-St. 4, 06120 Halle, Germany }

\pacs{73.23.-b}{Electronic transport in mesoscopic systems}
\pacs{73.23.Ra}{Persistent currents}
\pacs{42.65.Re}{Ultrafast processes;
optical pulse generation and pulse compression}
\abstract{We study theoretically the charge polarization and the
charge current dynamics of a mesoscopic ring driven by  short
asymmetric electromagnetic
 pulses and  threaded
 by an external static magnetic flux. It is shown that the
 pulse-induced charge polarization and the associated light-emission
 is controllable  by tuning the  external magnetic flux. Applying two
 mutually perpendicular pulses
 triggers a charge current  in the ring. The interplay between
 this nonequilibrium and the persistent currents is investigated
 and the conditions under which  the pulses stop the persistent current are
 identified.}

\begin{document}
 \maketitle

\section{INTRODUCTION}
\label{intro}  
Quantum structures with a ring confining geometry have served over the years as
a paradigm for the demonstration of quantum-interference  phenomena such as the
Aharonov-Bohm effect and the persistent (equilibrium) current
\cite{ring-theo1,ring-theo2,ring-exp,micromat} that emerges if the ring is
pierced by a static magnetic flux. The persistent current has also been
considered when  a time-dependent magnetic field with a static component
\cite{8n,9n,10n,11n,124,125,127,128} being applied, and it has been found that
the current diminishes when the static component vanishes. Another possibility
for generating (nonequilibrium) current is to irradiate  the ring with circular
polarized light, \cite{129,Pershin2005,Barth2006} and references therein. In
principle, this approach is  feasible for a size-quantized system and if the
radiation frequency is tuned such that the  rotating-wave approximation is
applicable
 and counter-rotating contributions to the current are thus negligible.
  Hence, this approach is expected to be
particularly useful for molecular ring structure, cf.
ref.~\cite{Barth2006} and references therein. Another approach for
current generation which does not rely on the level quantization
and resonant excitations, is based on an asymmetry of the electric
field  amplitude of the applied pulse \cite{idync,revivals}.  As
experimentally demonstrated \cite{hcp,seqhcp1,seqhcp2,seqhcp3} the
optical cycle of such asymmetric pulses consists of a short and
strong half cycle followed by a much longer and weak half cycle of
an opposite polarity. The charge dynamics is mainly triggered by
the strong , short half cycle
\cite{idync,revivals,AndPLA,Alex_PRB2004} and hence these pulses
are referred to  as half-cycle pulses (HCP). In general,  HCPs are
capable of inducing nonequilibrium  electric dipole and direct
charge currents \cite{revivals,AndPLA,Alex_PRB2004}, even if the
system is not  lacking  an inversion symmetry, in contrast to the
well-known photovoltaic effect \cite{60,61}. For isolated rings
(and without an external magnetic flux), applying one linearly
polarized HCP induces a charge polarization. Applying in a
perpendicular direction to the first pulse a second  HCP leads to
a charge current that can be tuned by varying  the properties and
the time-delay between the two pulses. The question which has not
yet been addressed and will be of concern here is how the
HCPs-induced charge polarization and the charge current are
influenced by the persistent current generated by  a static
magnetic flux. As shown below, the HCP-induced nonequilibrium
charge polarization and the associated emitted radiation are
tunable by changing the magnitude of the static magnetic flux. The
nonequilibrium current triggered by two time-delayed HCPs can be
tuned to cancel the magnetic-flux-induced equilibrium current
offering thus a possibility for time switching of the persistent
current on a picosecond time-scale.

\section{Theoretical model}
\label{theory}

For the sake of simplicity, we consider here an isolated single
channel 1D ballistic ring with $N$ electrons and radius $\rho_{0}$
at low temperatures. This treatment  is appropriate if the  width
and the height of the ring are smaller than both the ring radius
and the Fermi wavelength of the carriers. The generalization to
the multi-channel case can be done along the lines of
Refs.\cite{idync,alexepl} where it is shown that the multi-channel
case does not change qualitatively the predictions of the 1D
model. The ring is pierced by an external magnetic flux $\phi$ and
is subjected to a sequence of short, linearly polarized half-cycle
pulses. The light propagation direction is perpendicular to the
plane of the ring. The Hamiltonian describing the dynamics of the
system is given by
\begin{equation}\label{hamilt}
    \hat{H}(t)=\hat{H}_{0}+\hat{V}_{int},
\end{equation}
where
\begin{equation}\label{h0}
    \hat{H}_{0}=\frac{1}{2m^{\ast}\rho_{0}^{2}}\left(-i\hbar \frac{\partial}{\partial \theta}+\hbar
    \frac{\phi}{\phi_{0}}\right)^{2}\;\;;\;\;\hat{V}_{int}=e\mathbf{E}(t)\cdot\mbox{\boldmath$\rho$}.
\end{equation}
Here $e$ is the elementary charge, $\mathbf{E}(t)$ stands for the
time-dependent electric field of the applied pulses and the vector
$\mbox{\boldmath$\rho$}=(\rho_0\cos\theta,\rho_0\sin\theta)$ describes the
electron position.  $m^{\ast}$ and $\phi_{0}$ are the electron effective mass
and the flux quantum, respectively. In what follows we expand the field
operator $\hat{\Psi}(\theta,t)$
 on
a  basis of eigenstates $\varphi_n(\theta)$ of $\hat{H}_{0}$,
i.e.,
\begin{equation}\label{states}
\hat{\Psi}(\theta,t)=\sum_{n}\hat{c}_{n}(t)\varphi_n(\theta),\quad
\hat{H}_{0}\varphi_n(\theta) = E_{n}^{(0)}\varphi_n(\theta).
\end{equation}
 The components of the time-dependent density matrix are
expressed as $\rho_{mn}(t)=\langle
\hat{c}_{m}^{\dag}(t)\hat{c}_{n}(t)\rangle$. Here $\langle ...
\rangle$ denotes the expectation value which is taken with respect
to the initial state of the system.

Now we turn to the specification of the employed light pulses. The
duration $\tau_{d}$ of the experimentally available HCPs is in the
subpicosecond time scale.  On the other hand, for typical
ballistic mesoscopic rings ($\rho_0\sim 1 \mu m$) the time
$\tau_{_{F}}$ a particle at the Fermi level needs for a round trip
is tens of picoseconds, meaning that  $\tau_{d}\ll\tau_{_{F}}$. In
this case, the so-called  impulsive approximation (IA) can be
employed to  describe accurately the dynamics of the system
\cite{ia}. Within the IA the action of the HCPs is encompassed in
  the matching conditions (\ref{match1}) for the density matrix at
the instances before and after  the pulses are applied while the
system propagates in a pulse-free manner at any other
time\cite{note1} (note that in  absence of the pulses  the
magnetic flux is still present). To be specific, let us consider
the case where two HCPs that are linearly polarized in the $x$ and
$y$ directions are applied respectively at the times $t_{1}=0$ and
$t_{2}=\tau$. For the density matrix we find within the IA that
  (we use the abbreviation $|\varphi_n(\theta)\rangle \equiv
|n\rangle$)
\begin{equation}\label{match1}
  \begin{split}
    &\rho_{mn}(0^{+})=\sum_{k,l}\langle n|e^{-i\alpha_{1}\cos \theta} |l\rangle
     \langle k|e^{i\alpha_{1}\cos \theta} |m\rangle
     \rho_{kl}(0^{-});\\
     &\rho_{mn}(\tau^{+})=\sum_{k,l}\langle n|e^{-i\alpha_{2}\sin \theta} |l\rangle
     \langle k|e^{i\alpha_{2}\sin \theta} |m\rangle
     \rho_{kl}(\tau^{-}).
  \end{split}
\end{equation}
Here the dimensionless quantity  $\alpha_j$ characterizes the
action transferred to the system
 \begin{equation}\label{alpha}
 \alpha_{j}=\rho_{0}p_{j}/\hbar,
  \end{equation}
  while $p_j$ describes the strength of the
pulse $E_j$
  \begin{equation}\label{p_alpha}
 p_{j}=e\int_{t_{j}}^{t_{j}+\tau_d}
E_{j}(t)dt.
 \end{equation}
   The subindex $j=1,2$ refers to the
first and second pulse (with the same duration $\tau_d$). We remark here that
for a different propagation direction of HCPs, e.g. in the plane of the ring,
and/or for the strong excitation regime ($\alpha \gg 1$) the time-dependent
magnetic-field component of the pulse may in general affect the dynamics of the
electron system and should be included in the Hamiltonian \eqref{hamilt}. This
is to be contrasted with the case discussed in this paper where the magnetic
field associated with the HCP is moderate and lies in the plane of the ring
(where electrons are confined).

Within the relaxation time approximation the equation of motion
for the density matrix reads
\begin{equation}\label{dmatrix}
  \begin{split}
    \frac{\partial \rho_{mn}}{\partial
    t}=&i\frac{E_{m}^{(0)}-E_{n}^{(0)}}{\hbar}\rho_{mn}\\
    &-\frac{\delta_{mn}}{T_{1}}\left[\rho_{mn}-\rho_{nn}(0^{-})\right]
    -\frac{1-\delta_{mn}}{T_{2}}\rho_{mn},
  \end{split}
\end{equation}
where $\delta_{mn}$ is the Kronecker symbol and $T_{1}$ and
$T_{2}$ are the relaxation and dephasing times, respectively.
Eqs.(\ref{match1})-(\ref{dmatrix}) together with the initial
condition that before the application of the pulses the system is
in a thermal equilibrium, i.e.,
\begin{equation}\label{dinit}
    \rho_{mn}(0^{-})=
    2\delta_{mn}\left(1+e^{-\frac{[E_{n}^{(0)}-\eta]}{k_{_{B}}T}}\right)^{-1},
\end{equation}
 determine completely the time evolution of the density matrix. The
factor 2 in the Fermi-Dirac distribution (\ref{dinit}) accounts
for the two-fold spin degeneracy and the chemical potential $\eta$
is set  by the requirement  of the conservation of the  particles'
number $N$.

We are particularly interested in the study of the charge
polarization and the currents induced in the ring. The charge
polarization along the $x$ axis is determined by
the $x$-component of the dipole moment operator.
After some algebra one obtains for
the charge polarization
\begin{equation}\label{pol}
    \mu(t)
    =-e\rho_{0}\textrm{Re}\left[\sum_{n}\rho_{n,n-1}(t)\right].
\end{equation}
This result is remarkable. It tells us that the time dependence of
the charge polarization does not depend on the time evolution of
the diagonal components of the density matrix, i.e., by tracing
the evolution of the induced polarization one can obtain physical
information that is exclusively due to dephasing.

Using the first of eqs.~(\ref{match1}) and eqs.~(\ref{dmatrix})-(\ref{pol}) and
performing the same steps as in ref.~\cite{Alex_PRB2004} we can write the
charge polarization of the ring after the application of the first HCP as
\begin{equation}\label{Eq:mu_general}
  \begin{split}
    \mu(t)=-&e\rho_0\alpha_1\Theta(t)[J_0(\Omega)+J_2(\Omega)]\sin\!\!\left[\frac{2\pi t}{t_p}\right]
    e^{-t/T_2}\\
            &\times\sum_n \rho_{nn}(0^-)\cos\!\!\left[\frac{4\pi t}{t_p}\left(n+\frac{\phi}{\phi_0}\right)\right],
  \end{split}
\end{equation}
where $\Theta(x)$ denotes the Heaviside step function and $J_l(x)$
is a Bessel function of the order $l$. Furthermore, we introduced
the quantities
\begin{equation}\label{Eq:Omega}
    \Omega=\alpha_1\sqrt{2-2\cos(4\pi t/t_p)},
\end{equation}
\begin{equation}\label{Eq:t_p}
    t_p=4\pi m^{\ast}\rho_{0}^{2}/\hbar.
\end{equation}

The charge current density operator can be expressed in terms of
the field operators (\ref{states})
and the angular component of the vector potential
$A_{\theta}=\phi/(2\pi \rho_{0})$ as follows
\begin{equation}\label{jop}
  \begin{split}
        \hat{j}=& \frac{ie\hbar}{2m^{\ast}\rho_{0}^{2}}
    \left[\hat{\Psi}^{\dag}(\theta,t)\frac{\partial \hat{\Psi}(\theta,t)}{\partial \theta}-
    \frac{\partial \hat{\Psi}^{\dag}(\theta,t)}{\partial
    \theta}\hat{\Psi}(\theta,t)\right]\\
    & -\frac{e^2}{\rho_{0}m^{\ast}c}\hat{\Psi}^{\dag}(\theta,t)A_{\theta}\hat{\Psi}(\theta,t).
  \end{split}
\end{equation}
Upon  mathematical manipulations we find  for the charge current
\begin{equation}\label{ic}
I(t)=\langle \langle \hat{j}\rangle \rangle =
    I_{0}\sum_{n}(n+\phi/\phi_{0})\rho_{nn}(t).
\end{equation}
Here
\begin{equation}\label{inull}
I_{0}=-e\hbar/(m^{\ast}\rho_{0}^{2})
\end{equation}
 sets the scale
of the current magnitude. The double bracket $\langle \langle
\ldots \rangle \rangle$ stands for angular integration and
expectation value computation. In contrast to the charge polarization, the
charge current depends only on the diagonal elements of the density matrix.
Therefore, dephasing and relaxation can be independently investigated by
tracing the charge polarization and the charge current, respectively.

\begin{figure*}[t]
  \centering
  \epsfig{file=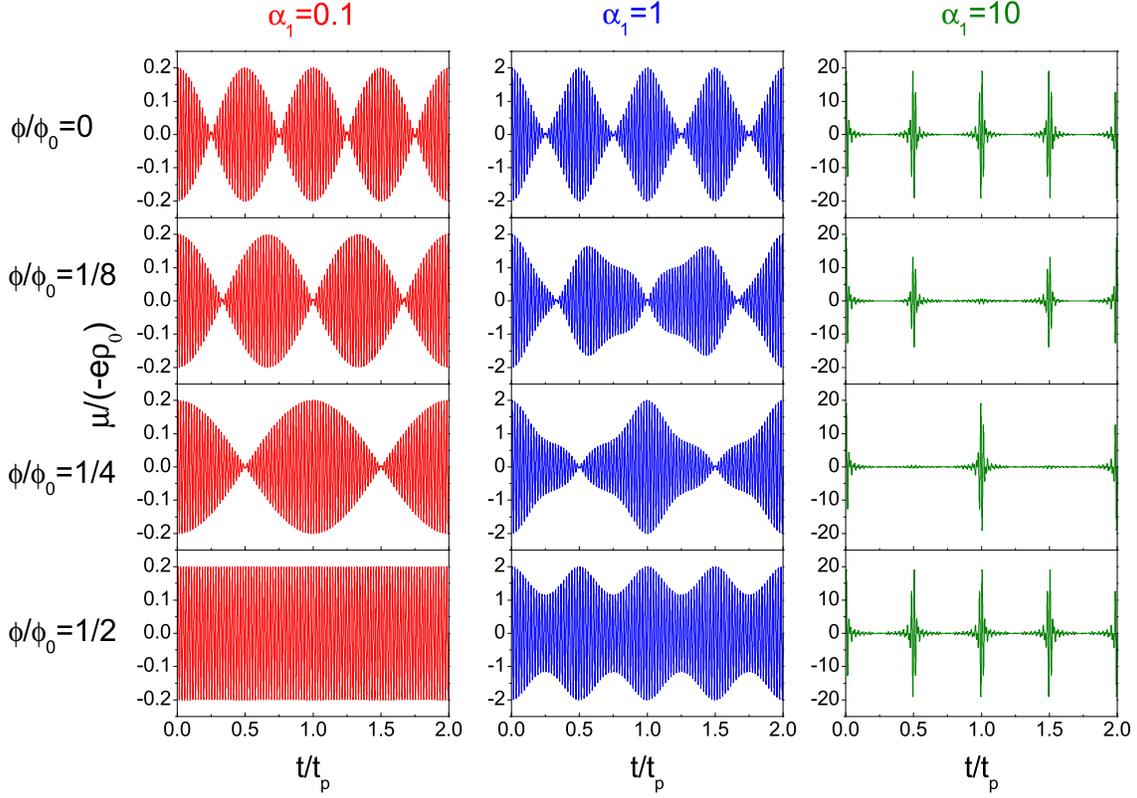,width=15.0cm}
   \caption{The time dependence of the charge polarization $\mu (t)$
   as a function of the
    magnetic flux $\phi$ for various values of the pulse strengths that result
    in
   different
   values of the transferred action $\alpha_{1}$ (cf.Eq.(\ref{alpha})). For the sin-square HCP having duration $\tau_d=1$~ps and
   the ring radius of $r_0=1~\mu$m the value
    of the transferred action $\alpha_1=0.1$ ($\alpha_1=1$, $\alpha=10$) corresponds to the peak electric field strength
    $E_0=1.32~$V/cm ($E_0=13.2~$V/cm, $E_0=132~$V/cm). The time is
   expressed in units of $t_{p}=4\pi m^{\ast} \rho_{0}^{2}/\hbar$.
   \label{F1a}}
\end{figure*}

Exploiting  Eqs.~(\ref{match1})-(\ref{dmatrix}) and (\ref{ic}) we
deduce after some algebra the following relation for the total
charge current\footnote{Self-inductive effects
turned out to be completely negligible in comparison
to $I_{pers}$ and $I_{dyn}(t)$.}
\begin{equation}\label{itot}
    I(t)=I_{pers}+\Theta(t-\tau)I_{dyn}(t),
\end{equation}
where
\begin{equation}\label{ipers}
    I_{pers}=I_{0}\sum_{n}(n+\phi/\phi_{0})\rho_{nn}(0^{-})
\end{equation}
is a static (persistent current) component induced by the magnetic
flux and
\begin{equation}\label{idyn}
    I_{dyn}(t)=\frac{I_{0}\alpha_{2}}{e\rho_{0}}\mu(\tau)e^{-t/T_{1}}
\end{equation}
is a dynamical component due to the action of the pulses and is
determined by the polarization $\mu(\tau)$ of the ring at the time
the second pulse is applied \cite{idync}. Note also that the
dynamical component of the current only appears after the second
pulse is applied, i.e., it is necessary to apply two orthogonal
linearly polarized HCPs in order to produce an additional
clock-wise anti-clock-wise symmetry breaking \cite{idync} (in case
of the persistent current this symmetry break is brought about by
the  magnetic-flux induced  time-reversal symmetry breaking).

It follows from eq.~(\ref{itot}) that the static part of the total charge
current induced in the ring does not depend on the application of the HCPs and
is not modified by them (i.e., the persistent current is really {\it
persistent}). On the other hand, the dynamical component $I_{dyn}$ depends on
both the external magnetic flux and the pulse parameters because, as shown
below explicitly, the dipole moment $\mu(\tau)$ is a function of $\phi$.

An important consequence of eqs.~(\ref{itot}) and (\ref{idyn}) is that a
magnetic-flux induced persistent current in a ballistic ring can
 be temporally stopped by applying HCPs, for all the parameters determining the
sign and the magnitude of $I_{dyn}$ are known [cf.
eqs.~(\ref{alpha}),~(\ref{inull}) and (\ref{Eq:mu})]. We note that the build-up
time of $I_{dyn}$ (which determines the switch-off time of $I$, i.e. the
cancellation of $I_{pers}$) is set by the duration of the HCP which can be in
the subpicosecond regime.  As $I_{dyn}$ relaxes according to eq.~(\ref{idyn})
$I_{pers}$ emerges again, however, upon the application of another sequence of
HCPs we can switch off again the total current (\ref{itot}).

It should be remarked here that the decomposition of the charge
current in a  statical and a dynamical component is valid within
the relaxation time approximation employed here. It might well be
that the inclusion of memory and non-linear effects prevents such
a simple structure of the current. A definitive conclusion on this
point would require  the treatment  of two-particle and higher
order correlations that is a challenging task for inhomogeneous
distributions (non-diagonal density matrix elements). On the other
hand, for weak excitations and small temperatures we expect
nonlinear effects (in the field strength and in relaxation) to be
marginal and the correlation effects on the relaxation to be of
less importance (due to the same reasons as for the ground state).

\section{Explicit results and numerical demonstrations}
\label{results}

For zero temperature we were able to perform the sum in the
eq.~\eqref{Eq:mu_general} analytically to obtain
\begin{equation}\label{Eq:mu}
    \mu(t)=-e\rho_0\alpha_1\Theta(t)[J_0(\Omega)+J_2(\Omega)]s(t)e^{-t/T_2},
\end{equation}
where
\begin{equation}\label{Eq:s_t}
  \begin{split}
    s(t)\!=&2\cos\!\left[\frac{4\pi t}{t_p}\left(\left|\frac{\phi}{\phi_0}\right|\!-\!\frac{1}{2}\right)\right]
    \sin\!\left[\frac{N\pi t}{t_p}\right],\hspace{0.2cm} N\!=0\ ({\rm
    mod}\; 4);\\
    s(t)\!=&\cos\!\left[\frac{4\pi t}{t_p}\frac{\phi}{\phi_0}\right]
    \!\sin\!\left[\frac{(N+1)\pi t}{t_p}\right]\\
         &+\cos\!\left[\frac{4\pi t}{t_p}\left(\left|\frac{\phi}{\phi_0}\right|\!-\!\frac{1}{2}\right)\right]
    \!\sin\!\left[\frac{(N-1)\pi t}{t_p}\right],\\
    &\hspace{5.5cm} N\!=1\ ({\rm
    mod}\; 4);\\
    s(t)\!=&2\cos\!\left[\frac{4\pi t}{t_p}\frac{\phi}{\phi_0}\right]
    \sin\!\left[\frac{N\pi t}{t_p}\right],\hspace{1.55cm} N\!=2\ ({\rm
    mod}\; 4);\\
    s(t)\!=&\cos\!\left[\frac{4\pi t}{t_p}\left(\left|\frac{\phi}{\phi_0}\right|\!-\!\frac{1}{2}\right)\right]
    \!\sin\!\left[\frac{(N+1)\pi t}{t_p}\right]\\
         &+\cos\!\left[\frac{4\pi t}{t_p}\frac{\phi}{\phi_0}\right]
    \!\sin\!\left[\frac{(N-1)\pi t}{t_p}\right],\hspace{0.55cm} N\!=3\ ({\rm
    mod}\; 4).
  \end{split}
\end{equation}
These results apply for $\phi/\phi_0\in[-1/2,1/2]$, otherwise
$s(t;\phi/\phi_0)$ is given by the periodicity condition
$s(t;\phi/\phi_0+1)$=$s(t;\phi/\phi_0)$. In the limiting case $\phi=0$
eqs.~\eqref{Eq:s_t} simplify to the results obtained in
ref.~\cite{Alex_PRB2004}. Inserting the expressions for $\mu(t)$ given by
eqs.~\eqref{Eq:mu} and \eqref{Eq:s_t} at $t=\tau$ into eq.~\eqref{idyn} yields
an analytical expression for the dynamical part of the current. The persistent
part of the current was calculated previously in ref.~\cite{Loss1991}.

For an illustration we performed explicit calculations for the case of a 1D
GaAs ring with $N=100$, radius $\rho_{0}=1\;\mu \textrm{m}$  at zero
temperature. The time dependence (ignoring the effects of dephasing
\cite{note2}) of the induced charge polarization is shown in fig.~\ref{F1a} for
different values of the external magnetic flux $\phi$ and pulse strengths
yielding different  $\alpha_{1}$ [cf. eq.~(\ref{alpha})]. The time is expressed
in units of the characteristic time $t_{p}$. For certainty we assume here that
the time profile of the HCP electric field is given by $E(t)=E_0\sin^2[\pi
t/\tau_d]$ for $t\in(0,\tau_d)$ and $E(t)=0$ outside of this time interval. In
this case we have $\alpha_1=\rho_0 e E_0\tau_d/(2\hbar)$. As seen from
fig.~\ref{F1a} different patterns of the charge polarization evolution can be
tailored by changing the applied flux and/or the pulse strength. This
observation is of particular relevance since the emission properties  of the
ring are determined by the time oscillations of the charge polarization
\cite{revivals,radiation}. Thus, the system proposed here could serve as a
source of electromagnetic radiation with magnetic-flux controllable properties.
Another remarkable fact revealed by fig.~\ref{F1a} is the presence of beating
behaviour of the  charge polarization dynamics in the low excitation regime as
well as collapses and revivals of the polarization at higher excitations
signifying the important role of  quantum interferences \cite{revivals}.

\begin{figure}[t]
   \centering
   \includegraphics[width=8.5cm]{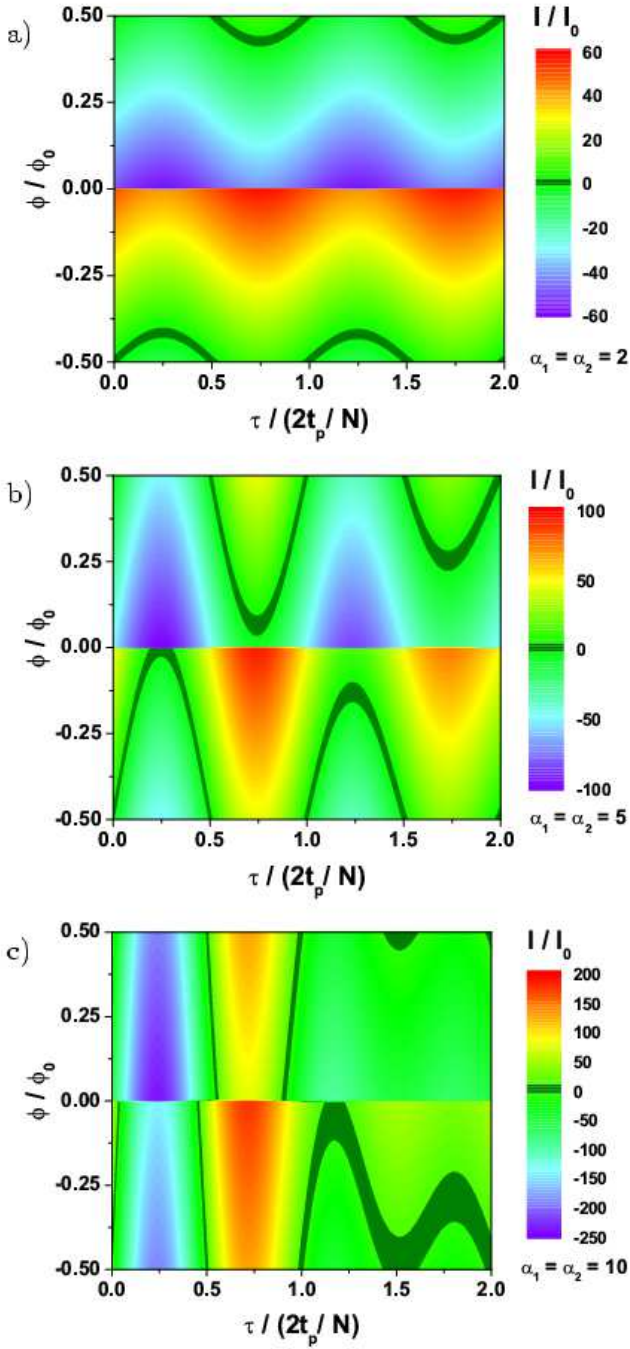}
   \caption
   { Dependence of the peak current $I$ on the magnetic flux $\phi$ and the time
   delay $\tau$ between the pulses for different values of the pulse strengths:
   a) $\alpha_1=\alpha_2=2$, b) $\alpha_1=\alpha_2=5$ and c) $\alpha_1=\alpha_2=10$. \label{Fig2}}
\end{figure}

Now we consider the current dynamics. As an example, fig.~\ref{Fig2} shows the
dependence of the peak current on the magnetic flux and the time delay between
the two pulses that trigger $I_{dyn}$. The peak current is given in units of
$I_{0}$ (e.g., for a ring with $\rho_{0}=1\;\mu \textrm{m}$ and $N=100$ we have
$I_{0}=3\textrm{ nA}$). The total charge current has an oscillatory behavior
when scanning $\phi$ and $\tau$. In particular, both the magnitude and the sign
of the peak current can be changed by appropriately varying $\phi$ and $\tau$.
As we outlined above, there exist regions in which the total induced current
vanishes due cancellation  of the persistent and the dynamical components of
the current. As it is seen from the figure the region where the total current
is switched off is controllable by sweeping  $\phi$, $\tau$ and/or the pulse
strengths. All of these parameters are externally tunable and experimentally
feasible with current technology.

Strictly speaking, the derived analytical formulas,
(\ref{itot})-(\ref{idyn}) and ((\ref{Eq:mu}-\ref{Eq:s_t})), are
only justified for the weak excitation regime. In the strong
excitation regime the nonlinear terms in relaxation can couple
diagonal and non-diagonal components of the density matrix, and in
this way the dynamics of the polarization and current. However, if
the relaxation is slow (and we can expect it to be so for low
temperatures) it hardly  affects the results presented in the
numerical illustrations also for the strong excitation regime
($\alpha=5,10$).

Finally we note that the circulating current results in an induced
(orbital) magnetization ($M(t) \approx \pi \rho_{0}^{2}I(t)$). The
current temporal controllability is reflected in a picosecond
switching of the ring magnetization.  The above finding can be
generalized, particularly in view of  potential applications,  to
the case of trains of HCPs pulses and/or arrays of mesoscopic
rings \cite{revivals,idync}. The first case may serve as a
sophisticated source of ultrashort magnetic pulses  while the
second  allows the  design of  artificially microstructured
materials with magnetic properties locally controllable.  In view
of applying the present scheme to nano and molecular rings
\cite{Barth2006} we note that this regime is accessed by changing
appropriately the parameters of HCPs: The applicability of the
above theory is based on the assumption  $\tau_d\ll t_p$ which
implies  the use of femtosecond HCP for nano-size rings. Such
pulses and even attosecond HCPs are currently under discussions
\cite{hcp_burg}.

\section{Conclusions}

In summary, we showed that the time-dependent charge polarization and charge
currents can be generated in mesoscopic rings threaded by a magnetic flux and
subjected to a sequence of asymmetric electromagnetic pulses. The time
dependence of the induced polarization shows different patterns that can be
controlled by appropriately adjusting the magnetic flux and the pulse
parameters. Quantum interferences result in  beatings and revivals of the
charge polarization dynamics. The total charge current induced in the ring can
be decomposed into a static and a dynamical components. The static component is
associated with the persistent current and  does not depend on the parameters
of the pulses.  The dynamical component is expressible analytically and can be
tuned to cancel the persistent current allowing thus a picosecond switching of
the total current in the ring.
%

%


\begin{thebibliography}{1}
\bibitem{note1} Note that we consider time delays $\tau$ between
 consecutive pulses that are much shorter than the relaxation and
 dephasing times. Relaxation and dephasing during the time interval $t<\tau$ is then
 neglected.
\bibitem{note2} Within the relaxation time approximation, the
 effect of dephasing consists in an exponential decay of the polarization on a
 time scale of the order of $T_{2}$. For the system considered here $T_{2}$ is
 several tens of nanoseconds \cite{revivals}.
\end{thebibliography}
\end{document}